\documentclass[twocolumn,amsmath,amssymb,prd,reprint,longbibliography,floatfix,8pt]{revtex4-1}

\usepackage{cancel}
\usepackage{enumitem}
\usepackage[utf8x]{inputenc}
\usepackage{graphicx}
\usepackage{dcolumn}
\usepackage{bm}
\usepackage[switch]{lineno}
\usepackage{float} 
\usepackage{subfigure}
\usepackage{tikz}
\usepackage{multirow}
\usetikzlibrary{arrows.meta}
\usetikzlibrary{arrows,decorations.pathmorphing,backgrounds,positioning,fit,petri}

\begin{document}

\title{The spin-phonon relaxation mechanism of single-molecule magnets in the presence of strong exchange coupling}

\author{Sourav Mondal$^1$}
\author{Julia Netz$^2$}
\author{David Hunger$^3$}
\author{Simon Suhr$^4$}
\author{Biprajit Sarkar$^{4,5}$}
\author{Joris van Slageren$^3$}
\author{Andreas K\"ohn$^2$}
\author{Alessandro Lunghi$^1$}
\email{lunghia@tcd.ie}
\affiliation{$^1$ School of Physics, AMBER and CRANN Institute, Trinity College, Dublin 2, Ireland}
\affiliation{$^2$ Institute for Theoretical Chemistry, University of Stuttgart, Pfaffenwaldring 55, D-70569 Stuttgart, Germany}
\affiliation{$^3$ Institute of Physical Chemistry, University of Stuttgart, Pfaffenwaldring 55, D-70569 Stuttgart, Germany}
\affiliation{$^4$ Institute of Inorganic Chemistry, University of Stuttgart, Pfaffenwaldring 55, D-70569 Stuttgart, Germany}
\affiliation{$^5$ Institute for Chemistry and Biochemistry, Freie Universit\"at Berlin, Fabeckstraße 34-36, 14195 Berlin, Germany}

\begin{abstract}
{\bf Magnetic relaxation in coordination compounds is largely dominated by the interaction of the spin with phonons. Large zero-field splitting and exchange coupling values have been empirically found to strongly suppress spin relaxation and have been used as the main guideline for designing new molecular compounds. Although a comprehensive understanding of spin-phonon relaxation has been achieved for mononuclear complexes, only a qualitative picture is available for polynuclear compounds. Here we fill this critical knowledge gap by providing a full first-principle description of spin-phonon relaxation in an air-stable Co(II) dimer with both large single-ion anisotropy and exchange coupling. Simulations reproduce the experimental relaxation data with excellent accuracy and provide a microscopic understanding of Orbach and Raman relaxation pathways and their dependency on exchange coupling, zero-field splitting, and molecular vibrations. Theory and numerical simulations show that increasing cluster nuclearity to just four cobalt units would lead to a complete suppression of Raman relaxation. These results hold a general validity for single-molecule magnets, providing a deeper understanding of their relaxation and revised strategies for their improvement.}
\end{abstract}

\maketitle

\section*{Introduction}

Coordination complexes with large magnetic anisotropies have been shown to display slow relaxation of their magnetic moment. Compounds of this class have been named single-molecule magnets (SMMs), in recognition of their similarities with hard magnets\cite{sessoli1993magnetic}, and have attracted a large interest for potential applications such as nanomagnetism\cite{gatteschi2006molecular}, spintronics\cite{sanvito2011molecular} and more recently quantum information science\cite{atzori2019second}. 

One of the main priorities for advancing these applications is mitigating the dramatic increase of the spin relaxation rate with temperature, $T$. The spin dynamics of a molecular spin is deeply affected by its interaction with the surrounding lattice, namely the spin-phonon coupling, which ultimately leads to a state with a zero expectation value of the magnetization\cite{lunghi2023spin}. At high temperatures, the spin relaxation time, $\tau$, follows the Arrhenius law
\begin{equation}
 \tau=\tau_0 e^{U_\mathrm{eff}/\mathrm{k_B} T}    \:,
 \label{Arr}
\end{equation}
where the effective magnetization reversal barrier $U_\mathrm{eff}$ can be taken as a measure of magnetic anisotropy\cite{perfetti2019multiple}. Very early on, it was recognized that in the absence of under-barrier processes, the effective energy barrier is $U_\mathrm{eff}=|D|S^2$, where $D$ is the molecular axial zero-field splitting and $S$ is the ground-state total spin value. For negative values of $D$, the maximum values of the spin projections, $M_S=\pm S$, are the lowest in energy, and up to $2S$ sequential phonon quanta have to be absorbed/emitted to go from $M_S=S \rightarrow M_S=-S$ and reverse the spin orientation. As the factor $|D|S^2$ increases in value, phonons with high energy (and thus low thermal population) are needed to initiate this process, leading to a slower relaxation, as expressed by Eq. \ref{Arr}. This relaxation mechanism falls under the name of Orbach and maximising $D$ has proved to be the best strategy to reduce its detrimental effects on the spin\cite{neese2011not,rinehart2011exploiting}. To this end, much effort has been devoted to the design of mononuclear coordination compounds and outstanding results have been achieved, e.g. $U_\mathrm{eff}\sim$ 450 cm$^{-1}$\cite{bunting2018linear} and 1500 cm$^{-1}$\cite{guo2018magnetic} in linear Co(II) and Dy(III) mononuclear complexes, respectively. 

However, minimizing the effect of Orbach relaxation is not the sole challenge in the field, and other spin relaxation mechanisms are operative. For instance, Quantum Tunneling of the Magnetization (QTM) operates at low temperatures and is primarily responsible for the closure of the magnetic hysteresis curve at magnetic zero field. Exchange coupling was recognized to slow down QTM and its study has been at the centre stage of the field since the very beginning\cite{wernsdorfer2002exchange}. 

However, the maximization of both zero-field splitting and exchange coupling, in the attempt to quench both Orbach and QTM relaxation, has proved challenging\cite{swain2023strategies}. On the one hand, as the field moved to SMMs with large single-ion zero-field splittings, the phenomenology of spin-phonon relaxation has become much more complex and Raman and direct phonon-mediated relaxation mechanisms have started to play a role at low temperature and high magnetic field, respectively\cite{zadrozny2011slow,rechkemmer2016four}. Moreover, whilst early SMMs, like Mn$_{12}$\cite{sessoli1993magnetic}, and Fe$_4$\cite{accorsi2006tuning}, were mostly based on ions with low zero-field splitting and large exchange coupling values, most compounds with large zero-field splitting fall in a weak or intermediate exchange coupling regime\cite{demir2015radical}. Molecules of this class present a multitude of low-energy spin states available for relaxation, making the interpretation of experiments far from trivial and unequivocal\cite{pedersen2015design,rechkemmer2016four}. 

In recent years, the combination of ab initio computational methods and the theory of open quantum systems has made it possible to simulate spin-phonon relaxation in magnetic molecules\cite{lunghi2022toward,lunghi2023spin}, opening a window on the microscopic details of this process. Ab initio simulations of spin relaxation have played a key role in resolving conflictual interpretations of spin relaxation in mononuclear compounds, particularly in the interpretation of Raman relaxation\cite{lunghi2020limit,briganti2021complete}, and resolving a large number of misinterpretations of experiments\cite{mondal2022unraveling}. A first-principles description of spin-phonon coupling and relaxation in SMMs has not only aided the interpretation of the experiments\cite{moseley2018spin,mondal2022unraveling} but has also provided insights into the nature of spin-phonon coupling and its relation to molecular structure\cite{lunghi2017intra,reta2021ab,dey2024unravelling}. However, ab initio studies of spin relaxation have so far been exclusively performed for mononuclear compounds and our understanding of spin relaxation in polynuclear SMMs is still largely phenomenological. In order for the field of exchange-coupled SMMs to progress it is now imperative to achieve a complete understanding of their spin relaxation and address the urgent questions of the nature of spin-phonon coupling and relaxation mechanisms in these compounds. 

In this contribution, we use ab initio open quantum systems theory to address spin relaxation in polynuclear clusters and complete the microscopic interpretation of Orbach and Raman spin-phonon relaxation in SMMs. For this purpose, we use the radical-bridged Co(II) dimer [K(18-crown-6]$_2$][K(H$_2$O)$_4$][Co(bmsab)]$_2$($\mu$-tmsab)] ($\mathbf{Co_2Rad}$), where bmsab is the dianion of 1,2-bis(methanesulfonamide) benzene and tmsab is the radical-trianion of 1,2,4,5-tetrakis(methanesulfonamide)benzene. This recently reported compound\cite{hunger2024electronic} exhibits some of the largest $U_\mathrm{eff}$ and slowest relaxation among transition metal complexes. Importantly, unlike most Ln-based SMMs, this compound is air-stable, potentially representing an optimal building block for future technologies based on SMMs. Last but not least, the electronic structure of this compound has recently been exhaustively characterized\cite{hunger2024electronic}, and spin relaxation in its mononuclear building block\cite{rechkemmer2016four} has been successfully explained with ab initio methods\cite{lunghi2020multiple}, thus offering an ideal starting point for our study.
\begin{figure}[h!]
    \centering
    \includegraphics[scale=1]{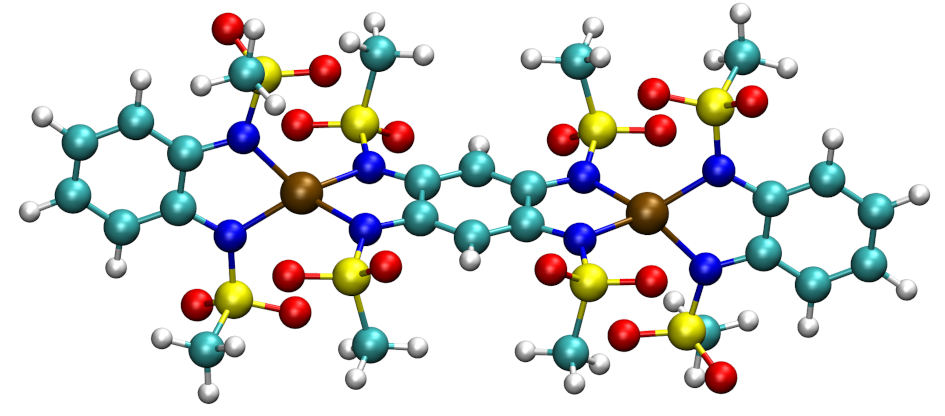}
    \caption{\textbf{$\mathbf{Co_2Rad}$ molecular structure.} The three-dimensional structure of the anion $\mathbf{Co_2Rad}$ is reported with the colour code: Co in brown, H in white, C in green, S in yellow, O in red, and N in blue.}
    \label{Figgeo}
\end{figure}

\section*{Methods}

{\bf Geometry optimization and phonons.} Cell and geometry optimization and simulations of $\Gamma$-point phonons are performed with periodic Density Functional Theory (DFT) using the CP2K software\cite{kuhne2020cp2k}. Cell optimization are performed with a very tight force convergence criterion of 10$^{-7}$ a.u. and SCF convergence criteria of 10$^{-10}$ a.u. for the energy. A plane wave cutoff of 1000 Ry, DZVP-MOLOPT Gaussian basis sets, and Goedecker-Tetter-Hutter Pseudopotentials\cite{goedecker1996separable} are employed for all atoms. The Perdew-Burke-Ernzerhof (PBE) functional and DFT-D3 dispersion corrections are used \cite{perdew1996generalized,grimme2010consistent}. Phonons are computed with a two-step numerical differentiation of atomic forces using a step of 0.01 \AA. 

\textbf{Electronic excitations and spin Hamiltonian.} The electronic structure of the dinuclear Co(II) complex is computed with complete active space self-consistent field (CASSCF) and multistate CAS perturbation theory to second order (MS-CASPT2) using the Molpro progam package\cite{MOLPRO_brief,MOLPRO_JCP_2020}.
Computations use a CAS(19,14) (19 electrons distributed in 14 orbitals) comprised of the two 3d orbital sets and 4$\pi$ orbitals of the radical bridge unit.
The molecular orbitals are represented by def2-TZVPP basis sets\cite{Weigend2005} for Co and N, and def2-SVP basis sets\cite{Weigend2005} otherwise, along with the appropriate auxiliary basis sets for density fitting\cite{Weigend2005}.
The orbitals are optimized with 4 octet states, and 8 sextet, 8 quartet, and 8 double states included in the state-averaging and the combined first-order/second-order algorithm of Kreplin and co-workers was employed\cite{Kreplin2020}.
For the MS-CASPT2 computations, the local pair-natural-orbital (PNO) based implementation of Werner and co-workers is used\cite{Menezes2016,Kats2019}, 
along with a shift of 0.45 E$_\text{h}$.
The detailed input parameters are given in the ESI.
For the lowest states of each multiplet, the multistate corrections are negligible and the runs for determining the numerical derivatives of the exchange coupling are done without multistate corrections.

The $g$ and zero-field splitting tensors of the Co(II) centres are determined in separate computations, where one centre is diamagnetically substituted with Zn(II) and the bridging ligand is considered in its reduced diamagnetic state. Here, a CAS(7,5) is employed, which comprised the 3d orbitals of Co(II).
Configuration-averaged Hartree-Fock orbitals (a state-averaging over all possible states weighted by their multiplicity)\cite{McWeeny1974} is used and all possible CAS configuration interaction (CASCI) states are determined (10 quartet and 40 doublet states).
The spin-orbit Hamiltonian is set up for these CASCI state and augmented with the CASPT2 correlation energies and off-diagonal MS-CASPT2 coupling matrix elements.
The Breit-Pauli spin-orbit Hamiltonian and the one-centre approximation are used.
The pseudospin formalism is employed to extract the $g$ and zero-field splitting tensors\cite{Chibotaru2012}.

{\bf Spin-phonon coupling.} Spin-phonon coupling coefficients $(\partial \hat{H}_{s}/\partial Q_{\alpha})$ are computed as
\begin{equation}
    \left( \frac{\partial \hat{H}_{s}}{\partial Q_{{\alpha}}} \right )= \sum_{i}^{3N} \sqrt{\frac{\hbar}{2\omega_{\alpha}m_{i}}} L_{\alpha i}\left( \frac{\partial \hat{H}_{s}}{\partial X_i} \right )\:.
    \label{num_diff}
\end{equation}
where $Q_{{\alpha}}$ is the dimensionless displacement vector associated with the phonon mode $\alpha$ and $N$ is the number of atoms in the unit cell. $L_{\alpha i}$ and $\omega_{\alpha}$ are the corresponding eigenvectors of the Hessian and the angular frequency. Only $\Gamma$-point phonons are considered. The first-order derivatives of the spin Hamiltonian with respect to the Cartesian coordinate $X_i$, $(\partial \hat{H}_{s} / \partial X_i)$, is computed by numerical differentiation. Each molecular degree of freedom is sampled four times between $\pm$ 0.1 \AA$ $. \\

\textbf{Spin-phonon relaxation.} Once the eigenstates, $| a \rangle$, and eigenvalues, $E_{a}$, of the spin Hamiltonian, $\hat{H}_s$, have been obtained, the spin dynamics can be simulated by computing the transition rate among different spin states, $W_{ba}$. Spin relaxation in Kramers systems with large magnetic anisotropy consists of contributions from one- and two-phonon processes. Considering one-phonon processes, the transition rate, $\hat{W}^{1-ph}_{ba}$, between spin states reads\cite{lunghi2017role,lunghi2019phonons} 
\begin{equation}
    \hat{W}^{1-ph}_{ba}=\frac{2\pi}{\hbar^2} \sum_{\alpha} | \langle b| \left(\frac{\partial \hat{H}_{s}}{\partial Q_{\alpha}} \right) |a \rangle  |^2G^{1-ph}(\omega_{ba}, \omega_{\alpha}) \:, 
    \label{Orbach}
\end{equation}
where $\hbar\omega_{ba}=E_{b}-E_{a}$. The function $G^{1-ph}$ reads
\begin{equation}
G^{1-ph}(\omega, \omega_{\alpha}) = \delta(\omega-\omega_\alpha)\bar{n}_\alpha +\delta(\omega +\omega_\alpha)(\bar{n}_\alpha +1)    \:,
\label{G1}
\end{equation}
where $ \bar{n}_\alpha=(e^{\hbar\omega_{\alpha}/\mathrm{k_B}T} -1)^{-1}$ is the Bose-Einstein distribution accounting for the thermal population of phonons, $\mathrm{k_B}$ is the Boltzmann constant, and the Dirac delta functions, e.g. $\delta(\omega-\omega_\alpha)$, enforce energy conservation during the absorption and emission of phonons by the spin system, respectively. Eq. \ref{Orbach} accounts for the Orbach relaxation mechanism\cite{lunghi2023spin}, where a series of phonon absorption processes leads the spin from the fully polarized state $M_{s}=S$ to an excited state with an intermediate value of $M_{s}$ before the spin can emit phonons back to $M_{s}=-S$. 

Two-phonon transitions, $W^{2-ph}_{ba}$, are responsible for Raman relaxation and include the absorption of two phonons, emission of two phonons or absorption of one phonon and emission of a second one. The latter process is the one that determines the Raman relaxation rate at low temperatures and is modelled as\cite{lunghi2022toward}
\begin{equation}
    \hat{W}^{2-ph}_{ba}  =\frac{2\pi}{\hbar^2} \sum_{\alpha\beta}\left | T^{\alpha\beta,+}_{ba} + T^{\beta\alpha,-}_{ba} \right|^2G^{2-ph} (\omega_{ba}, \omega_{\alpha}, \omega_{\beta})\:,
    \label{Raman}
\end{equation}
where the terms
\begin{equation}
T^{\alpha\beta,\pm}_{ba} = \sum_{c} \frac{ \langle b| (\partial \hat{H}_{s}/\partial Q_{{\alpha}}) |c\rangle \langle c| (\partial \hat{H}_{s}/\partial Q_{{\beta}})|a\rangle }{E_c -E_a \pm \hbar\omega_\beta} 
\label{Raman2}
\end{equation}
involve the contribution of all the spin states $|c\rangle $ at the same time, often referred to as a virtual state. The function $G^{2-ph}$ fulfils a similar role as $G^{1-ph}$ for one-phonon processes and it includes contributions from the Bose-Einstein distribution and imposes energy conservation. For the absorption/emission of two phonons, $G^{2-ph}$ reads
\begin{equation}
G^{2-ph}(\omega, \omega_{\alpha},\omega_{\beta}) = \delta(\omega-\omega_\alpha+\omega_\beta)\hat{n}_\alpha(\hat{n}_\beta +1).
\label{G2}
\end{equation}
All two-phonon contributions to $\hat{W}^{2-ph}_{ba}$ are included and their full equations are reported elsewhere\cite{lunghi2022toward}. All the parameters appearing in Eqs.~\ref{Orbach} and ~\ref{Raman} are computed from first principles. The Dirac delta functions appearing in Eqs. \ref{G1} and \ref{G2} are replaced by Gaussians with a smearing of 15 cm$^{-1}$. As discussed elsewhere, this substitution is a good approximation for a vanishing smearing and a full sampling of the phonons' Brillouin zone and corresponds to treating the bath as harmonic\cite{lunghi2020multiple,lunghi2020limit,lunghi2022toward}. Tests on the consistency of predictions of the spin relaxation time $\tau$ with respect to the Gaussian smearing are reported in ESI. 

The simulation of Kramers systems in zero external fields requires the use of the non-diagonal secular approximation\cite{lunghi2022toward}, where population and coherence terms of the density matrix are not independent of one another. This is achieved by simulating the dynamics of the entire density matrix for one-phonon processes. The full expression of Eq. \ref{Orbach} is reported in the literature\cite{lunghi2019phonons,lunghi2022toward}. On the other hand, an equation that accounts for the dynamics of the entire density matrix under the effect of two-phonon processes resulting from fourth-order time-dependent perturbation theory is not yet available. However, it is possible to remove the coupling between population and coherence terms by orienting the molecular easy axis along the quantization $z$-axis and by applying a small magnetic field to break Kramers degeneracy\cite{lunghi2022toward}. Here we employ the latter strategy to simulate Raman relaxation. 

Once all the matrix elements $W^{n-ph}_{ba}$ have been computed, $\tau^{-1}$ can be predicted by simply diagonalizing $W^{n-ph}_{ba}$ and taking the eigenvalue corresponding to an eigenvector describing a population transfer between the states of the ground-state KD. This usually corresponds to the smallest non-zero eigenvalue. The software MolForge\cite{lunghi2022toward}, freely available at github.com/LunghiGroup/MolForge, is used for all these simulations.\\

\section*{Results}

\textbf{Electronic structure and magnetic interactions.} The electronic structure of $\mathbf{Co_2Rad}$ is determined with multireference computational methods as described in Methods. The spin-free energy levels obtained from MS-CASPT2 calculations are reported in Table \ref{tab:spin_ladder} and fitted to a Heisenberg Hamiltonian of the form
\begin{equation}
    \hat{H}_H=  J (\:\vec{\mathbf{s}}_1  \cdot \vec{\mathbf{s}}_b 
              + \:\vec{\mathbf{s}}_2  \cdot \vec{\mathbf{s}}_b ) \:,
    \label{SH_Heisenberg}
\end{equation}
where $\vec{\mathbf{s}}_1$, $\vec{\mathbf{s}}_2$, and $\vec{\mathbf{s}}_b$ are the pseudospin operators of the two Co(II) centres and the radical bridge, respectively. Due to the inversion symmetry of $\mathbf{Co_2Rad}$, the exchange coupling, $J$, is identical for the two Cobalt ions. 
\begin{table}[htb]
    \caption{Spin ladder of $\mathbf{Co_2Rad}$: $\Delta E$ are the relative MS-CASPT2 energies of the states in the absence of spin-orbit coupling. The levels are assigned to the spin-ladder states of the Heisenberg Hamiltonian, Eq.~\ref{SH_Heisenberg}, resulting in effective values for the $J$ parameter, $J_\text{eff}$.}
    \label{tab:spin_ladder}
    \centering
    \begin{ruledtabular}
    \begin{tabular}{dccd}
\multicolumn{1}{c}{$\Delta E$ / cm$^{-1}$} &	$S$	& assignment &	
\multicolumn{1}{c}{$J_\text{eff}$/ cm$^{-1}$}\\
\hline
   0.0 & 5/2 &	0 J   &	 \\
 198.1 &	3/2	& 1/2 J & 396.3 \\
 394.5 &	1/2	& J	    & 394.5 \\
 783.9 &	5/2	&	    & \\
 783.7 &	5/2	&	    & \\
 919.2 &	1/2	& 2 J	& 459.6 \\
 973.8 &	3/2	&	    & \\
 974.1 &	3/2	&	    & \\
1119.1 & 3/2 & 5/2 J & 447.6 \\
1165.6 & 1/2 &		 & \\
1167.1 & 1/2 &		 & \\
1319.9 & 5/2 & 3 J	 & 440.0 \\
1383.6 & 7/2 & 7/2 J & 395.3 \\
    \end{tabular}
    \end{ruledtabular}
\end{table}

In accordance with previous findings\cite{hunger2024electronic}, the present computations confirm a ferri-magnetic coupling of the two Co(II) centres via the radical bridge and a ground-state with total spin $S=5/2$ and $J$ values in the range 395 to 440~cm$^{-1}$. As the lowest multiplets usually dominate the magnetic properties and spin dynamics at low temperatures, we set $J=396.3$ cm$^{-1}$, corresponding to the value for the spin Hamiltonian model from the lowest energy gap, between the sextet and the quartet. Note that the inclusion of dynamic correlation is vital for accurate exchange couplings, the $J$ value extracted from CASCI states would only be 169~cm$^{-1}$. The solution of the spin Hamiltonian with $J=396.3$ cm$^{-1}$ gives a spin ladder with total spin $S = 5/2$, $3/2$, $1/2$, $1/2$, $3/2$, $5/2$, $7/2$, which nicely fits the states up to energy $\sim$ 784~cm$^{-1}$. At higher energies, states originate from excited quartet states of the Co(II) centres and give rise to further spin ladders\cite{hunger2024electronic}, as indicated in Tab.~\ref{tab:spin_ladder}.

The effect of spin-orbit coupling is included through the zero-field splitting Hamiltonian
\begin{equation}
    \hat{H}_{ZFS}= \vec{\mathbf{s}}_1 \cdot \mathbf{D} \cdot \vec{\mathbf{s}}_1 
             + \vec{\mathbf{s}}_2 \cdot \mathbf{D} \cdot \vec{\mathbf{s}}_2              
    \label{SH}
\end{equation}
where $\mathbf D$ represents the zero-field splitting tensor of the two Co(II) ions. As described in Methods, the projection to a pseudospin Hamiltonian for a single quartet state results in a highly axial zero-field splitting tensor with $D = -114.1$~cm$^{-1}$ and $E = 0.9$~cm$^{-1}$. The main anisotropy axis is directed along the long axis of the molecule, connecting the two Co(II) centres. A previous detailed spectroscopy study\cite{hunger2024electronic} resulted in $J = 390$~cm$^{-1}$ and $D = -113$~cm$^{-1}$ (with $E$ close to zero), confirming the accuracy of the present parametrization for $J$ and  $D$. Finally, in order to simulate the effect of a static magnetic field we also compute the Zeeman Hamiltonian
\begin{equation}
    \hat{H}_Z = \mu_B \vec{\mathbf{s}}_1 \cdot \mathbf{g} \cdot \vec{\mathbf{B}} 
             + \mu_B \vec{\mathbf{s}}_2 \cdot \mathbf{g} \cdot \vec{\mathbf{B}}
             + \mu_B \vec{\mathbf{s}}_b \cdot \mathbf{g}_b \cdot \vec{\mathbf{B}}  
    \label{SHZ}
\end{equation}
The values of the g-tensors, $\mathbf{g}$, are also extracted from ab initio simulations, while $\mathbf{g}_b$ is set to the free-electron value. The total spin Hamiltonian is the sum of the three terms in Eqs. \ref{SH_Heisenberg}-\ref{SHZ} and the numerical values of all its tensors are provided in ESI. The eigenvalues of the total spin Hamiltonian are reported in the ESI, where it can be seen that the system features a well-separated sextet with considerable zero-field splitting as the lowest state and a maximum $M_S$ quantum number for the ground-state Kramers doublet (KD).

\begin{figure}[h!]
    \centering
    \includegraphics[scale=1]{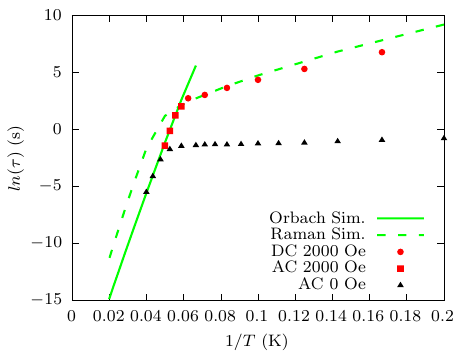}
    \caption{\textbf{Spin relaxation time.} Computed Orbach and Raman relaxation times are reported with continuous and dashed green lines, respectively. Red square and circle symbols correspond to experimental relaxation times extracted from AC and DC magnetometry, respectively, in the presence of a magnetic field of 2000 Oe. The black triangles report the experimental relaxation time measured with AC magnetometry in zero external field.}
    \label{Fig1}
\end{figure}
\textbf{Spin-Phonon Relaxation.} The effect of phonons on spin dynamics can be treated within the framework of perturbation theory. In particular, the linear coupling terms of Eq. \ref{num_diff} are known to influence the dynamics of spin at both the second and fourth order of perturbation theory\cite{lunghi2023spin}. Fig. \ref{Fig1} reports the comparison between simulated and experimental spin relaxation times\cite{hunger2024electronic}. In the absence of an external magnetic field, the experimental relaxation time for $\mathbf{Co_2Rad}$ is found to be almost insensitive to temperature up to $\sim$~20~K. Above this threshold, a stark reduction is observed, coherently with reports in similar compounds\cite{albold2019strong}. The relaxation mechanism is interpreted as driven by QTM and spin-phonon coupling in the two temperature regimes, respectively. The comparison with simulations shows good agreement only for the highest temperature measurement, but the two diverge otherwise. Here we are interested in understanding the role of phonons and we thus perform a comparison with experiments obtained in the presence of an external magnetic field\cite{hunger2024electronic} to quench the effect of QTM and uncover the intrinsic limits to spin lifetime imposed by spin-phonon relaxation. AC magnetometry results show that relaxation time increases and an exponential regime is revealed for increasing field values. For $T$ lower than $\sim$~15~K, relaxation time becomes too long to be tracked by AC magnetometry, and DC is instead used. Under the applied field the agreement between simulations and experimental values is excellent and clearly shows that AC tracks the Orbach regime, whilst the simulated Raman relaxation well describes slower DC values.

\begin{figure}[b!]
    \centering
    \begin{tikzpicture}

\def\linelength{0.5}
\def\rescale{99.4989}

\draw[black, thick] ( -2.4999251242244078 , 0.0000000000000000 / \rescale) -- (\linelength+ -2.4999251242244078 , 0.0000000000000000 / \rescale);
\draw[black, thick] ( 2.4999246591004010 , 1.1889996305175146 / \rescale) -- (\linelength+ 2.4999246591004010 , 1.1889996305175146 / \rescale);
\draw[black, thick] ( -1.4981704122619601 , 233.46487387554680 / \rescale) -- (\linelength+ -1.4981704122619601 , 233.46487387554680 / \rescale);
\draw[black, thick] ( 1.4981168440603649 , 234.17947282676482 / \rescale) -- (\linelength+ 1.4981168440603649 , 234.17947282676482 / \rescale);
\draw[black, thick] ( -0.49818901646272751 , 299.37390700799233 / \rescale) -- (\linelength+ -0.49818901646272751 , 299.37390700799233 / \rescale);
\draw[black, thick] ( 0.49824303578892959 , 299.60685162403320 / \rescale) -- (\linelength+ 0.49824303578892959 , 299.60685162403320 / \rescale);
\draw[black, thick] ( -0.49967839792353691 , 362.54401474736028 / \rescale) -- (\linelength+ -0.49967839792353691 , 362.54401474736028 / \rescale);
\draw[black, thick] ( 0.49966017778070171 , 362.77311964084458 / \rescale) -- (\linelength+ 0.49966017778070171 , 362.77311964084458 / \rescale);
\draw[black, thick] ( -1.4996399780775884 , 397.27115165465614 / \rescale) -- (\linelength+ -1.4996399780775884 , 397.27115165465614 / \rescale);
\draw[black, thick] ( 1.4996581360583408 , 397.99543135161582 / \rescale) -- (\linelength+ 1.4996581360583408 , 397.99543135161582 / \rescale);
\draw[black, thick, -{Stealth[scale=0.75]}] ( -2.99993 ,-0.25) -- ( 3.49992 ,-0.25);
\draw[black, thick, -{Stealth[scale=0.75]}] ( -2.99993 ,-0.25) -- ( -2.99993 , 397.99543135161582 / \rescale+0.5);
\draw[black, thick, -{Stealth[scale=0.75]}] ( -2.99993 ,-0.25) -- ( -2.99993 , 397.99543135161582 / \rescale+0.5);
\node at ( -2.99993 -0.7, 1.2 / \rescale) { 0 };
\node at ( -2.99993 -0.7, 234.2 / \rescale) { 234 };
\node at ( -2.99993 -0.7, 299.6 / \rescale) { 300 };
\node at ( -2.99993 -0.7, 362.8 / \rescale) { 363 };
\node at ( -2.99993 -0.7, 398.0 / \rescale) { 398 };
\node at ( -2.5 +\linelength/2,-0.5) { -2.5 };
\node at ( 2.5 +\linelength/2,-0.5) { 2.5 };
\node at ( -1.5 +\linelength/2,-0.5) { -1.5 };
\node at ( 1.5 +\linelength/2,-0.5) { 1.5 };
\node at ( -0.5 +\linelength/2,-0.5) { -0.5 };
\node at ( 0.5 +\linelength/2,-0.5) { 0.5 };
\node at ( -0.5 +\linelength/2,-0.5) { -0.5 };
\node at ( 0.5 +\linelength/2,-0.5) { 0.5 };
\node at ( -1.5 +\linelength/2,-0.5) { -1.5 };
\node at ( 1.5 +\linelength/2,-0.5) { 1.5 };
\draw[black, thick] ( -2.99993 , 0.0000000000000000 / \rescale) -- ( -2.99993 +0.1, 0.0000000000000000 / \rescale );
\draw[black, thick] ( -2.99993 , 1.1889996305175146 / \rescale) -- ( -2.99993 +0.1, 1.1889996305175146 / \rescale );
\draw[black, thick] ( -2.99993 , 233.46487387554680 / \rescale) -- ( -2.99993 +0.1, 233.46487387554680 / \rescale );
\draw[black, thick] ( -2.99993 , 234.17947282676482 / \rescale) -- ( -2.99993 +0.1, 234.17947282676482 / \rescale );
\draw[black, thick] ( -2.99993 , 299.37390700799233 / \rescale) -- ( -2.99993 +0.1, 299.37390700799233 / \rescale );
\draw[black, thick] ( -2.99993 , 299.60685162403320 / \rescale) -- ( -2.99993 +0.1, 299.60685162403320 / \rescale );
\draw[black, thick] ( -2.99993 , 362.54401474736028 / \rescale) -- ( -2.99993 +0.1, 362.54401474736028 / \rescale );
\draw[black, thick] ( -2.99993 , 362.77311964084458 / \rescale) -- ( -2.99993 +0.1, 362.77311964084458 / \rescale );
\draw[black, thick] ( -2.99993 , 397.27115165465614 / \rescale) -- ( -2.99993 +0.1, 397.27115165465614 / \rescale );
\draw[black, thick] ( -2.99993 , 397.99543135161582 / \rescale) -- ( -2.99993 +0.1, 397.99543135161582 / \rescale );
\draw[black, thick] ( -2.4999251242244078 +\linelength/2,-0.25 ) -- ( -2.4999251242244078 +\linelength/2,-0.25+0.1);
\draw[black, thick] ( 2.4999246591004010 +\linelength/2,-0.25 ) -- ( 2.4999246591004010 +\linelength/2,-0.25+0.1);
\draw[black, thick] ( -1.4981704122619601 +\linelength/2,-0.25 ) -- ( -1.4981704122619601 +\linelength/2,-0.25+0.1);
\draw[black, thick] ( 1.4981168440603649 +\linelength/2,-0.25 ) -- ( 1.4981168440603649 +\linelength/2,-0.25+0.1);
\draw[black, thick] ( -0.49818901646272751 +\linelength/2,-0.25 ) -- ( -0.49818901646272751 +\linelength/2,-0.25+0.1);
\draw[black, thick] ( 0.49824303578892959 +\linelength/2,-0.25 ) -- ( 0.49824303578892959 +\linelength/2,-0.25+0.1);
\draw[black, thick] ( -0.49967839792353691 +\linelength/2,-0.25 ) -- ( -0.49967839792353691 +\linelength/2,-0.25+0.1);
\draw[black, thick] ( 0.49966017778070171 +\linelength/2,-0.25 ) -- ( 0.49966017778070171 +\linelength/2,-0.25+0.1);
\draw[black, thick] ( -1.4996399780775884 +\linelength/2,-0.25 ) -- ( -1.4996399780775884 +\linelength/2,-0.25+0.1);
\draw[black, thick] ( 1.4996581360583408 +\linelength/2,-0.25 ) -- ( 1.4996581360583408 +\linelength/2,-0.25+0.1);
\node at ( -2.99993 -0.8, 397.99543135161582 / \rescale+0.8) { $E$ (cm$^{-1}$) };
\node at ( 3.49992 +0.2 , -0.5 ) { $\langle S_z \rangle $ };
\draw[red, thick, -{Stealth[scale=0.75]}, opacity= 0.43 ] ( -2.4999251242244078 +\linelength/2, 0.0000000000000000 / \rescale) -- ( -1.4981704122619601 +\linelength/2, 233.46487387554680 / \rescale);
\draw[red, thick, -{Stealth[scale=0.75]}, opacity= 0.431017 ] ( 2.4999246591004010 +\linelength/2, 1.1889996305175146 / \rescale) -- ( 1.4981168440603649 +\linelength/2, 234.17947282676482 / \rescale);
\draw[red, thick, -{Stealth[scale=0.75]}, opacity= 0.0497961 ] ( -1.4981704122619601 +\linelength/2, 233.46487387554680 / \rescale) -- ( 1.4981168440603649 +\linelength/2, 234.17947282676482 / \rescale);
\draw[red, thick, -{Stealth[scale=0.75]}, opacity= 0.877334 ] ( -1.4981704122619601 +\linelength/2, 233.46487387554680 / \rescale) -- ( -0.49818901646272751 +\linelength/2, 299.37390700799233 / \rescale);
\draw[red, thick, -{Stealth[scale=0.75]}, opacity= 0.717783 ] ( -1.4981704122619601 +\linelength/2, 233.46487387554680 / \rescale) -- ( 0.49824303578892959 +\linelength/2, 299.60685162403320 / \rescale);
\draw[red, thick, -{Stealth[scale=0.75]}, opacity= 0.636708 ] ( -1.4981704122619601 +\linelength/2, 233.46487387554680 / \rescale) -- ( -0.49967839792353691 +\linelength/2, 362.54401474736028 / \rescale);
\draw[red, thick, -{Stealth[scale=0.75]}, opacity= 0.518912 ] ( -1.4981704122619601 +\linelength/2, 233.46487387554680 / \rescale) -- ( 0.49966017778070171 +\linelength/2, 362.77311964084458 / \rescale);
\draw[red, thick, -{Stealth[scale=0.75]}, opacity= 0.746592 ] ( -1.4981704122619601 +\linelength/2, 233.46487387554680 / \rescale) -- ( -1.4996399780775884 +\linelength/2, 397.27115165465614 / \rescale);
\draw[red, thick, -{Stealth[scale=0.75]}, opacity= 0.261179 ] ( -1.4981704122619601 +\linelength/2, 233.46487387554680 / \rescale) -- ( 1.4996581360583408 +\linelength/2, 397.99543135161582 / \rescale);
\draw[red, thick, -{Stealth[scale=0.75]}, opacity= 0.720621 ] ( 1.4981168440603649 +\linelength/2, 234.17947282676482 / \rescale) -- ( -0.49818901646272751 +\linelength/2, 299.37390700799233 / \rescale);
\draw[red, thick, -{Stealth[scale=0.75]}, opacity= 0.879508 ] ( 1.4981168440603649 +\linelength/2, 234.17947282676482 / \rescale) -- ( 0.49824303578892959 +\linelength/2, 299.60685162403320 / \rescale);
\draw[red, thick, -{Stealth[scale=0.75]}, opacity= 0.522789 ] ( 1.4981168440603649 +\linelength/2, 234.17947282676482 / \rescale) -- ( -0.49967839792353691 +\linelength/2, 362.54401474736028 / \rescale);
\draw[red, thick, -{Stealth[scale=0.75]}, opacity= 0.638386 ] ( 1.4981168440603649 +\linelength/2, 234.17947282676482 / \rescale) -- ( 0.49966017778070171 +\linelength/2, 362.77311964084458 / \rescale);
\draw[red, thick, -{Stealth[scale=0.75]}, opacity= 0.266874 ] ( 1.4981168440603649 +\linelength/2, 234.17947282676482 / \rescale) -- ( -1.4996399780775884 +\linelength/2, 397.27115165465614 / \rescale);
\draw[red, thick, -{Stealth[scale=0.75]}, opacity= 0.746562 ] ( 1.4981168440603649 +\linelength/2, 234.17947282676482 / \rescale) -- ( 1.4996581360583408 +\linelength/2, 397.99543135161582 / \rescale);
\draw[red, thick, -{Stealth[scale=0.75]}, opacity= 0.0295939 ] ( -0.49818901646272751 +\linelength/2, 299.37390700799233 / \rescale) -- ( 0.49824303578892959 +\linelength/2, 299.60685162403320 / \rescale);
\draw[red, thick, -{Stealth[scale=0.75]}, opacity= 0.947907 ] ( -0.49818901646272751 +\linelength/2, 299.37390700799233 / \rescale) -- ( -0.49967839792353691 +\linelength/2, 362.54401474736028 / \rescale);
\draw[red, thick, -{Stealth[scale=0.75]}, opacity= 0.997893 ] ( -0.49818901646272751 +\linelength/2, 299.37390700799233 / \rescale) -- ( 0.49966017778070171 +\linelength/2, 362.77311964084458 / \rescale);
\draw[red, thick, -{Stealth[scale=0.75]}, opacity= 0.819569 ] ( -0.49818901646272751 +\linelength/2, 299.37390700799233 / \rescale) -- ( -1.4996399780775884 +\linelength/2, 397.27115165465614 / \rescale);
\draw[red, thick, -{Stealth[scale=0.75]}, opacity= 0.590183 ] ( -0.49818901646272751 +\linelength/2, 299.37390700799233 / \rescale) -- ( 1.4996581360583408 +\linelength/2, 397.99543135161582 / \rescale);
\draw[red, thick, -{Stealth[scale=0.75]}, opacity= 1 ] ( 0.49824303578892959 +\linelength/2, 299.60685162403320 / \rescale) -- ( -0.49967839792353691 +\linelength/2, 362.54401474736028 / \rescale);
\draw[red, thick, -{Stealth[scale=0.75]}, opacity= 0.947897 ] ( 0.49824303578892959 +\linelength/2, 299.60685162403320 / \rescale) -- ( 0.49966017778070171 +\linelength/2, 362.77311964084458 / \rescale);
\draw[red, thick, -{Stealth[scale=0.75]}, opacity= 0.590753 ] ( 0.49824303578892959 +\linelength/2, 299.60685162403320 / \rescale) -- ( -1.4996399780775884 +\linelength/2, 397.27115165465614 / \rescale);
\draw[red, thick, -{Stealth[scale=0.75]}, opacity= 0.818156 ] ( 0.49824303578892959 +\linelength/2, 299.60685162403320 / \rescale) -- ( 1.4996581360583408 +\linelength/2, 397.99543135161582 / \rescale);
\draw[red, thick, -{Stealth[scale=0.75]}, opacity= 0 ] ( -0.49967839792353691 +\linelength/2, 362.54401474736028 / \rescale) -- ( 0.49966017778070171 +\linelength/2, 362.77311964084458 / \rescale);
\draw[red, thick, -{Stealth[scale=0.75]}, opacity= 0.882583 ] ( -0.49967839792353691 +\linelength/2, 362.54401474736028 / \rescale) -- ( -1.4996399780775884 +\linelength/2, 397.27115165465614 / \rescale);
\draw[red, thick, -{Stealth[scale=0.75]}, opacity= 0.639299 ] ( -0.49967839792353691 +\linelength/2, 362.54401474736028 / \rescale) -- ( 1.4996581360583408 +\linelength/2, 397.99543135161582 / \rescale);
\draw[red, thick, -{Stealth[scale=0.75]}, opacity= 0.640404 ] ( 0.49966017778070171 +\linelength/2, 362.77311964084458 / \rescale) -- ( -1.4996399780775884 +\linelength/2, 397.27115165465614 / \rescale);
\draw[red, thick, -{Stealth[scale=0.75]}, opacity= 0.882302 ] ( 0.49966017778070171 +\linelength/2, 362.77311964084458 / \rescale) -- ( 1.4996581360583408 +\linelength/2, 397.99543135161582 / \rescale);

\end{tikzpicture}

    \includegraphics[scale=1]{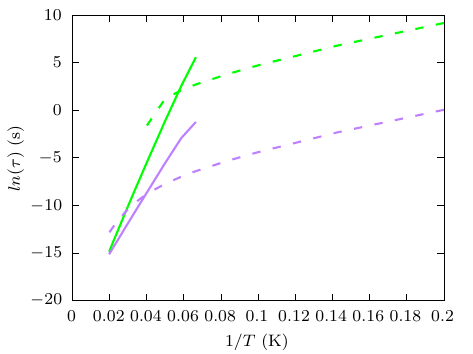}
    \caption{\textbf{Contributions to spin relaxation.} 
    Top panel: Computed transition rates due to one-phonon absorptions for the first five KDs. Only transitions with rates larger than 10$^{-12}$ ps$^{-1}$ have been considered. The color scale of the arrows is proportional to the logarithm of the rate. The $y$-axis reports the expectation value of the $z$ component of the total spin angular momentum. Bottom panel: Computed Orbach and Raman relaxation times are reported with continuous and dashed lines, respectively. Green curves report results for $\mathbf{Co_2Rad}$, and purple curves for $\mathbf{Co_1}$.}
    \label{Fig2}
\end{figure}
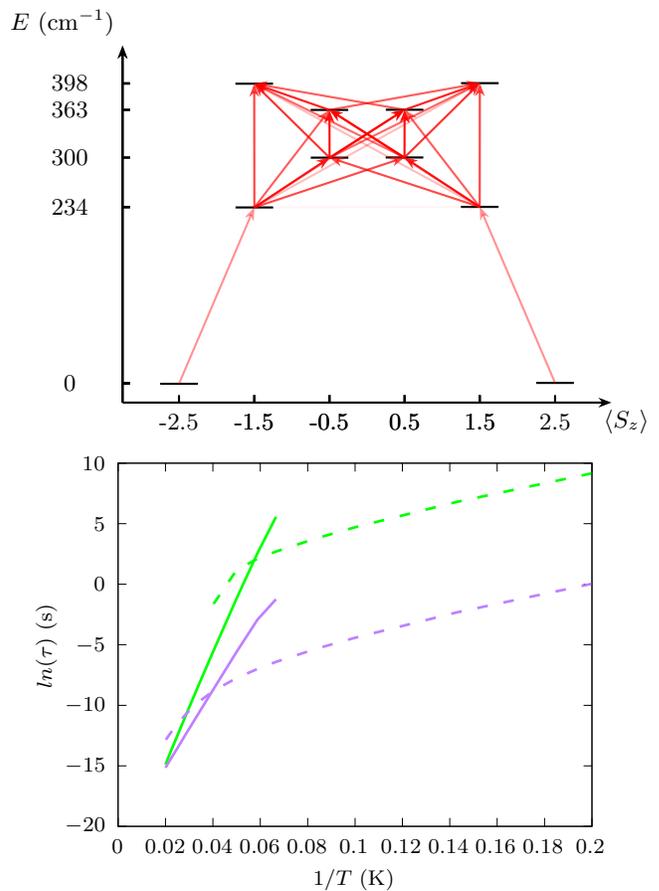

Aiming at interpreting the mechanism of relaxation for $\mathbf{Co_2Rad}$ in both Orbach and Raman regimes, we perform new simulations where only a part of the phonons are included. The slope of the $ln(\tau)$ vs. $1/T$ is canonically used to estimate the energy of the states involved in the Orbach process. This analysis returns a value of $\sim$ 300 cm$^{-1}$, thus commensurate with the energy of the second excited KD. This analysis, however, is not able to discern between a relaxation mechanism involving a single phonon with $\hbar\omega_\alpha \sim$ 300 cm$^{-1}$ inducing a single transition between the ground-state KD and the second excited KD, or the sequential absorption of two phonons whose energy add up to 300 cm$^{-1}$. In the latter case, a first phonon would induce a transition between ground-state KD and the first excited KD, while the second phonon would promote a subsequent transition from the first excited KD to the second excited KD. We observe that the computed Orbach relaxation rates drastically drop as soon as phonons with energy larger than $\sim 180$ are excluded from the simulation (see ESI), making it possible to assign the Orbach relaxation mechanism as promoted by two sequential absorption processes where phonons in resonance with the first excited KD $\sim$ 200 cm$^{-1}$ initiate the relaxation process. This can also be seen from the top panel of Fig. \ref{Fig2}, where the arrows point to the most probable transitions among different spin states. 

The study of the matrix element of Eq. \ref{Raman}, supports an interpretation of Raman relaxation at $T <$ 10 K as an intra-ground-state KD transition. In the giant spin representation of $\mathbf{Co_2Rad}$ this corresponds to a transition among states with $M_{S}= \pm 5/2$. We perform the simulations of the Raman relaxation rates by progressively removing phonons at high energy and identify the low-energy spectrum, below 50 cm$^{-1}$, to be the only relevant contribution (see ESI). We also perform the simulation of Raman rates at 7 K by progressively including more virtual transitions to KDs other than just the first excited one in the sum over $c$ in Eq. \ref{Raman2}. We note that the inclusion of the sole first KD produces a rate that is one order of magnitude faster, whilst the overall relaxation rate is well reproduced when the first two excited KDs are accounted for. However, the first five KDs are necessary to achieve full convergence (see ESI). These are interesting observations for two reasons: 1) the contributions to $\tau$ of different virtual transitions do not simply add up and cancellation effects can arise, and 2) states beyond the fundamental spin multiplet might play a role.

We now turn to the study of the contribution of different magnetic interactions to the spin-phonon relaxation mechanism. Firstly, we disentangle the contributions of single-ion anisotropy and exchange coupling on the overall molecular magnetic moments dynamics. In this attempt, we perform the simulations for a fictitious molecule $\mathbf{Co_1}$ where the magnetic moment on the radical and one of two Co ions have been quenched without affecting anything else. This is achieved by only considering the first term in Eq. \ref{SH} and its derivatives. Fig. \ref{Fig2} reports the comparison between spin relaxation in $\mathbf{Co_1}$ and $\mathbf{Co_2Rad}$, highlighting the large impact of exchange coupling in slowing down Orbach relaxation by up to two orders of magnitude and Raman relaxation by up to four orders of magnitude. A similar conclusion was reported by Albold et al. in comparing the mononuclear prototype [Co(L$_A$)$_2$]$^{2-}$ (H$_2$L$_A$ is 1,2-bis(methanesulfonamido)benzene) with the corresponding dimer\cite{albold2019strong} and precursor of  $\mathbf{Co_2Rad}$. Indeed the predicted dynamics of $\mathbf{Co_1}$ follows very closely the one of [Co(L$_A$)$_2$]$^{2-}$. 
\begin{figure}[h!]
    \centering
    \includegraphics[scale=1]{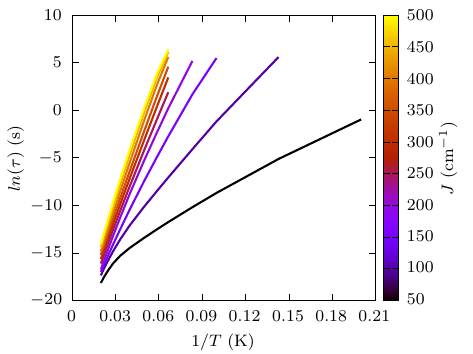} \\ 
    \includegraphics[scale=1]{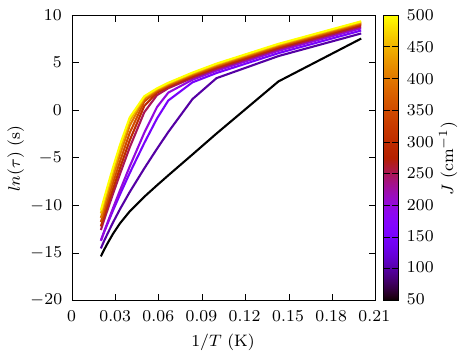} \\
    \includegraphics[scale=1]{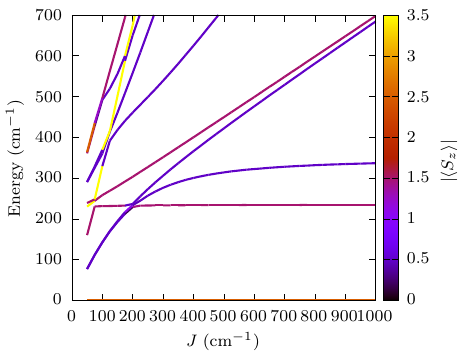}
    \caption{\textbf{Spin relaxation time vs. exchange coupling strength.} Computed Orbach (top panel) and Raman (middle panel) relaxation times are reported as a function of exchange coupling strength. The colour code for the lines is reported as a graph sidebar. The energy of the low-lying spin states is reported as a function of exchange coupling strength (bottom panel). The color code is reported as a graph sidebar and corresponds to the absolute value of the computed total spin $z$-component expectation value.}
    \label{Fig3}
\end{figure}

We now perform the simulation of spin-phonon relaxation as a function of the exchange coupling constant to isolate the importance of the coupling strength. As can be seen in Fig. \ref{Fig3}, both Orbach and Raman relaxation times rapidly increase with $J$ up to $\sim$ 200 cm$^{-1}$. For larger values of $J$, only marginal, but still visible, increments of relaxation times are observed. This can be explained by observing the bottom panel of Fig. \ref{Fig3}, where the lowest-energy states of the spin Hamiltonian are reported as a function of $J$ and colour-coded according to the expectation value of their total spin $z$-component, $\langle S_z \rangle$. Whilst the ground state always remains a pure $M_S=\pm 5/2$, the first few excited KDs drastically change their nature and only start settling for $J >$ 200 cm$^{-1}$. In this high-exchange regime, the molecule approximately behaves as a 5/2 giant spin and the first three KDs determine spin relaxation. The residual dependency of $\tau$ for $J >$ 400 cm$^{-1}$ signals that the giant-spin approximation is not perfectly fulfilled. For $J <$ 200 cm$^{-1}$, the presence of extra low-energy states offers additional relaxation pathways, cutting down the benefit of large single-ion zero-field splitting and even producing faster relaxation rates than in the mononuclear compound. The large rate of transition to states outside the fundamental multiplet is also visible in the top panel of Fig. \ref{Fig2}. Interestingly, we note that the contribution of exchange coupling is only operative at the level of the static Hamiltonian, therefore in shaping the spin wavefunction, but has no contribution at the level of spin-phonon coupling, as often invoked in classical literature\cite{abragam2012electron}. Indeed, if the derivatives of $J$ are omitted from Eq. \ref{num_diff}, results remain unchanged.

\begin{figure}[h!]
    \centering
    \includegraphics[scale=1]{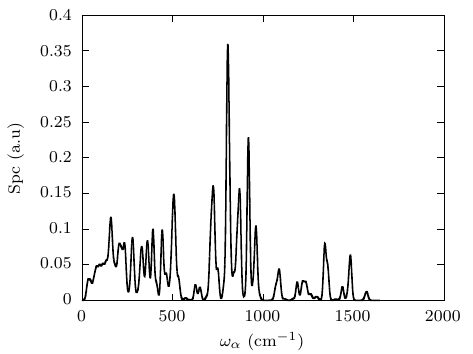} \\
    \includegraphics[scale=0.8]{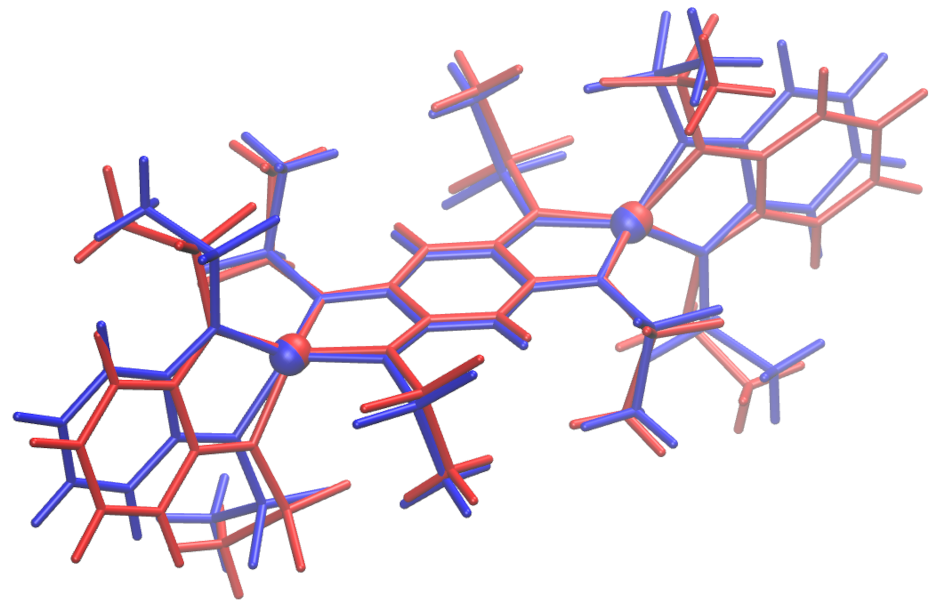} \\
    \includegraphics[scale=0.8]{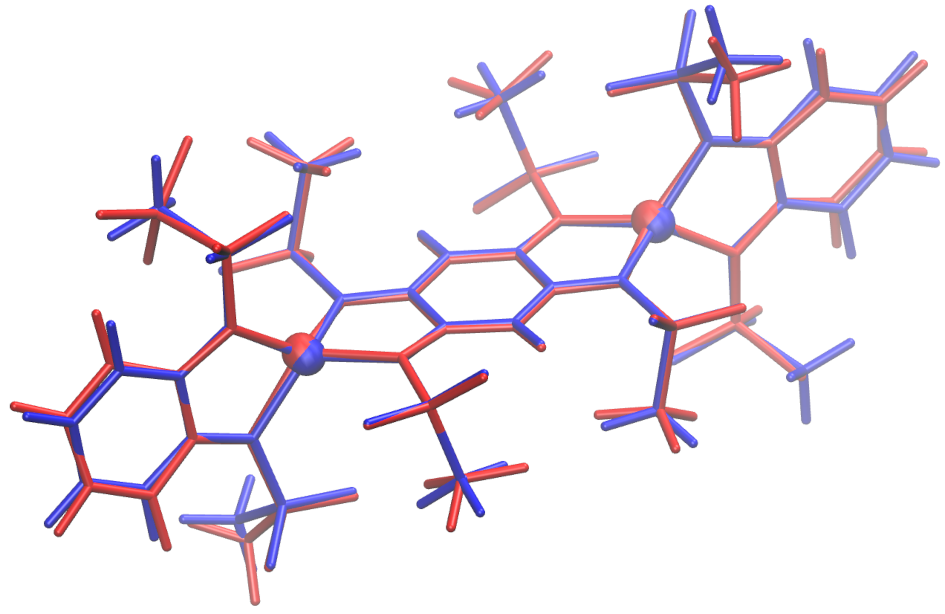}
    \caption{\textbf{Spin-phonon coupling distribution.} Top panel: The average spin-phonon coupling (Spc) of each phonon with spin is reported as their frequency. A Gaussian smearing of 10 cm$^{-1}$ is applied to smooth out the distribution. Middle panel: molecular distortions associated with the first available phonon at the $\Gamma$-point and responsible for Raman relaxation. The equilibrium molecule is reported in blue and the distorted one is in red. Bottom panel: molecular distortions associated with a phonon of energy $\sim$ 200 cm$^{-1}$ and responsible for the first step of Orbach relaxation. The equilibrium molecule is reported in blue and the distorted one is in red.}
    \label{Fig7}
\end{figure}

\clearpage
\textbf{Spin-Phonon Coupling.} Now that the relaxation mechanism has been determined we can turn our attention to the nature of spin-phonon coupling. We start by plotting how this quantity is distributed over the phonons density of state in Fig. \ref{Fig7}. 
As it is common for molecular crystals of this complexity, a continuous and highly structured distribution of phonons is present from very low wavenumber\cite{lunghi2019phonons,garlatti2023critical,mondal2022unraveling}, as also visible in Fig. \ref{Fig7} up to $\sim$ 250 cm$^{-1}$. A visual inspection of molecular displacements of phonons in this energy range shows atomic displacements characterized by complex motions delocalized over the entire unit cell. As we move to higher values of $\hbar\omega_\alpha$, the inter-molecular motions become less important and internal vibrations become more localized, also generating sharper peaks in the spin-phonon coupling distribution. Here we focus our attention on the two areas of this distribution that our previous analysis has found linked to spin dynamics, i.e. the lowest-energy phonons, responsible for Raman relaxation, and phonons around 200 cm$^{-1}$, responsible for the first step of the Orbach relaxation. In the former case, we observe motions where the aromatic rings of the ligands remain rigid and tilt with respect to one another (middle panel of Fig. \ref{Fig7}). On the other hand, the modes in resonance with the first KD (bottom panel of Fig. \ref{Fig7}) exhibit a large twisting of the aromatic rings of the bmsab ligands and methyl rotations. Admixed to these complex motions, Co-N bonds and the NCoN angles are also sensibly modulated, inducing a non-negligible spin-phonon coupling. 

\begin{figure}[h!]
    \centering
    \includegraphics[scale=1]{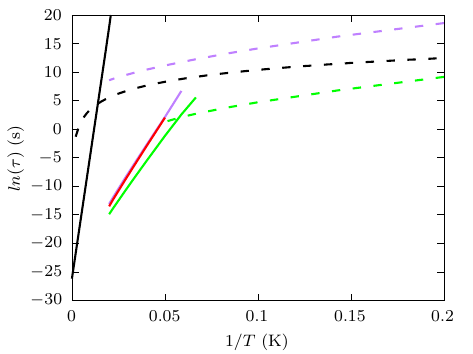}
    \caption{\textbf{Spin relaxation and nuclearity.} Orbach and Raman relaxation times are reported with continuous and dashed lines, respectively. Green lines are used for $\mathbf{Co_2Rad}$, purple lines for $\mathbf{Co_3}$ and red lines for $\mathbf{Co_4}$. Black lines report the fit of experimental Orbach and Raman relaxation times from ref. \cite{guo2018magnetic}. QTM contributions are neglected to make a direct comparison with the computed spin-phonon relaxation times.}
    \label{Fig6}
\end{figure}

\textbf{Relaxation in clusters of larger nuclearity.} Finally, we investigate the role of nuclearity in determining the relaxation time. To this end we simulate a hypothetical trimer, $\mathbf{Co_3Rad_2}$, where the original molecule is linearly extended, i.e. the first ion is exchange-coupled to the second one, and the latter is exchange-coupled to the third. We assume the spin Hamiltonian parameters and spin-phonon coupling to be identical to the dimer. Fig. \ref{Fig6} reports the simulated dynamics of $\mathbf{Co_3Rad_2}$ and shows that Raman relaxation slows down by four orders of magnitudes. Orbach relaxation also improves, though to a smaller degree, and relaxation at 20 K slows down by a factor of 20. Interestingly, if a tetramer, $\mathbf{Co_4Rad_3}$, is now considered, Raman relaxation times become so long that we cannot numerically estimate them, but Orbach relaxation remains virtually identical to $\mathbf{Co_3Rad_2}$. The latter result is consistent with the notion that $U_\mathrm{eff}$ is not dramatically affected by nuclearity alone as the total effective ZFS of the ground spin multiplet scales as $\sim 1/S^2$\cite{waldmann2007criterion,cirera2009build}, but it shows that Raman does not suffer from the same dependency and that it can be completely suppressed with a multi-ion strategy. Fig. \ref{Fig6} also compares these results with a state-of-the-art mononuclear Dy SMM\cite{guo2018magnetic}, showing that both $\mathbf{Co_3Rad_2}$ and $\mathbf{Co_4Rad_3}$ would support a slower Raman relaxation than this compound.

\section*{Discussion and Conclusions}

The study of $\mathbf{Co_2Rad}$ with ab initio methods has made it possible to shed light on important aspects of spin relaxation in polynuclear SMMs. Firstly, we have demonstrated that relaxation follows a similar trend to the mononuclear case, with Orbach and Raman relaxation mechanisms operative at high and low temperatures, respectively. This result provides a robust theoretical ground for the interpretation of relaxation experiments in polynuclear SMMs. For $\mathbf{Co_2Rad}$, we have shown that both relaxation mechanisms are largely well described by intra ground-state spin multiplet due to the large value of $J$ for this complex. However, this scenario would rapidly change as $J$ decreases below the value of the single-ion zero-field splitting. This analysis is in agreement with experiments that have observed a detrimental effect of coupling multiple ions\cite{rinehart2011strong,patrascu2017chimeric,liu2017single} and provides a quantitative framework for their interpretation. Importantly, while the effect of exchange on QTM has been known for a long time\cite{wernsdorfer2002exchange}, we have here been able to show that large exchange coupling has a dramatic effect on Raman relaxation. To the best of our knowledge, this behaviour has only been observed in experiments where a very large exchange coupling among magnetic centres is achieved\cite{rinehart2011strong,albold2019strong,gould2022ultrahard}, but never discussed in detail nor rationalized. These results also shed light on the limits of the giant-spin approximation to describe relaxation in SMMs. Our simulations show that electronic states beyond the fundamental $S=5/2$ multiplet play a role in Raman relaxation and to a smaller degree in Orbach relaxation. A similar situation has also been recently observed for the mononuclear case of cobalt ions, where excitations beyond the fundamental quartet have been shown to contribute to Raman virtual transitions\cite{mariano2023spin}.

It is worth stressing that the results achieved for $\mathbf{Co_2Rad}$ hold a general validity for polynuclear SMMs, including Ln-based ones. Indeed, whilst the chemistry of these compounds varies significantly, from a physical point of view their magnetic properties can be described within the same framework, i.e. zero-field splitting, exchange coupling and a thermal bath of molecular crystal vibrations. As it has been shown for the mononuclear case, these ingredients do not change qualitatively among different molecules and the same underlying principles of relaxation remain valid across the range of coordination compounds\cite{mondal2022unraveling}.

The compound investigated here is the result of years of molecular engineering and unsurprisingly our simulations show that both strong exchange coupling and large single-ion zero-field act in a concerted way to suppress spin relaxation. As a consequence, the question of how to achieve further progress in transition-metal-based SMMs beckons. Arguably, several strategies lie ahead: 1) further increase both single-ion zero-field splitting and exchange coupling, 2) engineer lattice and molecular vibrations, and 3) increase clusters' nuclearity.

Whilst the present molecule already exhibits some of the largest $U_\mathrm{eff}$ in transition-metal-based SMMs, further engineering of coordination complexes to increase $D$ and $J$ remains one of the most efficient ways to improve their performances. Similarly to what has been achieved with Dy ions, introducing a large exchange coupling among linearly coordinated cobalt ions\cite{bunting2018linear} might reveal unprecedentedly long relaxation times. In this regard, some of the present authors have presented a large-scale computational study of Co(II) coordination complexes\cite{mariano2024charting}, suggesting that novel coordination environments able to support maximally large values of axial zero-field splitting are yet to be explored and that the potential of transition-metal-based SMMs is still untapped. 

In terms of vibrational design, the visual inspection of the phonons of $\mathbf{Co_2Rad}$ shows a high degree of rigidity, with the key vibrations driving relaxation being already rather localized on the ligands and often involving the rotation of methyl groups. The latter motion has also been found to play a key role in the relaxation of Vanadyl compounds\cite{garlatti2023critical} and its substitution with a less flexible unit might represent a promising way to reduce low-energy vibrations. At the same time, we note that a consistent and quantitative definition of molecular rigidity and how it influences relaxation is yet to be achieved. We anticipate that additional theoretical efforts in this direction will be necessary to bring this design rule to fruition. 

Finally, we have explored how increasing the nuclearity of $\mathbf{Co_2Rad}$ could drastically influence its Raman relaxation mechanisms, slowing it down by several orders of magnitude. In particular, we have shown that whilst the upper limit for $U_\mathrm{eff}$ is achieved already at the level of three Co ions, Raman relaxation keeps scaling with nuclearity. This is a central result of our study and shows a clear way forward to controlling low-temperature relaxation. Given the chemical nature of this compound, we limited ourselves to the study of molecular chains, but the study could be expanded to more complex topologies. Interestingly, there is extensive literature on Co, or other ions, single-chain magnets using radical linkers\cite{caneschi2001cobalt,bernot2006family}. In this respect our analysis has two-fold importance: 1) our proposed synthetic guidelines to unprecedentedly slow Raman relaxations are likely to be well within the reach of synthetic chemistry, and 2) the scope of our simulations serves as a stepping stone toward the rationalization of decades of research in the study of relaxation in 1D magnetic systems and an ab initio description of their Glauber's dynamics. 

We must stress that all these considerations have been done without considering QTM, and therefore only remain valid in the presence of external field and/or magnetic dilution. The theoretical modelling of QTM is still in its infancy\cite{isakov2016understanding,aravena2018ab,mattioni2024vibronic} and we will need further development of relaxation theories to also account for this mechanism. We can however reasonably expect that as the effective spin ground state of SMMs increases with nuclearity, QTM should also be suppressed accordingly, provided no transverse zero-filed splitting is introduced.  

In conclusion, we have provided a full ab initio description of the spin relaxation mechanism in a paradigmatic air-stable SMM where both single-ion zero-field splitting and exchange coupling have been maximised. Importantly, simulations demonstrate that further extending the nuclearity of this compound from two to just three or four Co(II) ions could potentially lead to a compound with unprecedentedly slow Raman spin relaxation. These results hold a general validity for both 3d and 4f molecules and we anticipate that they will provide a new blueprint for the engineering of novel polynuclear SMMs and interpret their properties.

\vspace{0.2cm}
\noindent
\textbf{Acknowledgements and Funding}\\
A.L. and S.M. acknowledge support from the European Research Council (ERC) under the European Union’s Horizon 2020 research and innovation programme (grant agreement No. [948493]). A.K. and J.N. acknowledge support by the state of Baden-W\"{u}rttemberg through bwHPC and the German Research Foundation (DFG) through grant INST 40/575-1 FUGG (JUSTUS 2 cluster). J.v.S and D.H. acknowledge the support of the Landesgraduiert enforderung of the state of Baden-W\"{u}rttenberg (DFG SL103/10-2, SA1840-/9-1). A.L. and S.M. received computational resources by the Trinity College Research IT and the Irish Centre for High-End Computing (ICHEC). A.K. is also grateful for the support from the Stuttgart Center for Simulation Science (SimTech).

\vspace{0.2cm}
\noindent
\textbf{Conflict of interests}\\
The authors declare that they have no competing interests.


\end{document}